\documentclass[twocolumn,showpacs,amsmath,amssymb]{revtex4}
\usepackage{graphicx}
\usepackage{dcolumn}
\usepackage{bm}
\usepackage{amsmath}
\usepackage{amsfonts}
\begin{document}

\title{The Dark Side of Neutron Stars}
\author{Chris {\sc Kouvaris}}\email{kouvaris@cp3.sdu.dk}
\affiliation{$\text{CP}^3$-Origins, University of Southern Denmark, Campusvej 55, Odense 5230, Denmark}

\begin{abstract}
We review severe constraints on asymmetric bosonic dark matter based on observations of old neutron stars. Under certain conditions, dark matter particles in the form of asymmetric bosonic WIMPs can be effectively trapped onto nearby neutron stars, where they can rapidly thermalize and concentrate in the core of the star. If some conditions are met, the WIMP population can collapse gravitationally and form a black hole that can eventually destroy the star. Based on the existence of old nearby neutron stars, we can exclude certain classes of dark matter candidates. 
 \\[.1cm]
 {\footnotesize  \it Preprint: CP$^3$-Origins-2013-028 DNRF90 \& DIAS-2013-28.}\end{abstract}


\maketitle 

\section{Introduction}

Compact stars such as neutron stars and white dwarfs can lead in general to two types of constraints regarding dark matter candidates. The first one has to do with annihilating dark matter that changes the thermal evolution of the star. Annihilation of Weakly Interacting Massive Particles (WIMPs) that are trapped inside compact stars, can lead to the production of significant amount of heat that can change the temperature of old stars~\cite{Kouvaris:2007ay,Bertone:2007ae,Kouvaris:2010vv,deLavallaz:2010wp}.
 Such a phenomenon can be in principle contrasted to observations. The second type of constraints is related to asymmetric dark matter~\cite{Goldman:1989nd,Kouvaris:2011fi,McDermott:2011jp,Kouvaris:2011gb,Guver:2012ba,Fan:2012qy,Bell:2013xk,Jamison:2013yya}. Asymmetric dark matter is an attractive alternative to thermally produced dark matter especially due to the intriguing possibility of relating its asymmetry to the baryonic one. For recent reviews see~\cite{Petraki:2013wwa,Zurek:2013wia}. Due to the asymmetry, WIMP annihilation is not significant in this case. If a certain amount of WIMPs is trapped inside the star, the WIMPs can quite rapidly thermalize and concentrate within a tiny radius in the core of the star. If the WIMP population grows significantly, WIMPs might become self-gravitating and they might collapse forming a mini black hole. Under certain conditions, the black hole might consume the rest of the star, thus leading to the ultimate destruction of the star. However, very old (older than a few billion years) nearby neutron stars have been well observed and studied. The simple presence of such verified old stars leads to the conclusion that no black hole has consumed the star and as we shall argue, this can lead to very severe constraints on the properties of certain types of asymmetric dark matter. We should also mention that additional constraints on asymmetric dark matter can be imposed on different ways (e.g. from asteroseismology~\cite{Lopes:2012af,Casanellas:2012jp,Casanellas:2013nra}, from effects on the transport properties of the neutron stars~\cite{Horowitz:2012jd} and/or hybrid dark matter rich compact stars~\cite{Leung:2013pra,Goldman:2013qla}).

One can easily figure out that fermionic WIMPs due to the fact that they have to overcome Fermi pressure, require a huge number in order to collapse i.e. $N\sim (M_{pl}/m)^3$ where $M_{pl}$ and $m$ are the Planck mass and WIMP mass respectively. This number of WIMPs is very difficult to be accumulated within a few billion years and with dark matter densities similar to the ones of the earth. However, this required number for gravitational collapse is reduced significantly in the case of attractive Yukawa forces among the WIMPs~\cite{Kouvaris:2011gb}. 

\section{Asymmetric Bosonic Dark Matter}
In the case of asymmetric bosonic WIMPs, the necessary WIMP number for collapse is much smaller because there is no Fermi pressure and only the uncertainty principle keeps particles from collapsing. The collapse takes place once the momentum becomes smaller than the self-gravitational potential energy.
\begin{equation}
\frac{\hbar}{r}<\frac{GMm}{r} \Leftrightarrow M > \frac{M_{pl}^2}{m},
\end{equation}
where $M=Nm$ is the total mass of the WIMP cloud. A more accurate and generic estimate that includes the effect of self-interactions gives~\cite{Mielke:2000mh}
\begin{equation} 
M_{\rm crit}= \frac{2}{\pi} \frac{M_{pl}^2}{m} \sqrt{1+ \frac{\lambda M_{pl}^2}{32 \pi m^2}} \label{m_crit}. 
\end{equation}
  Although self-interactions between WIMPs can be quite general in nature, without loss of generality, we can assume that the self-interaction can be approximated well by a $\lambda \phi^4$ interaction term. At the no interaction limit $\lambda =0$ we trivially get the critical mass mentioned above (up to factors of order one).

 The accretion of WIMPs for a typical 1.4$M_{\bigodot}$ 10 km neutron star taking into account relativistic effects has been calculated in~\cite{Kouvaris:2010vv}. The total mass of WIMPs accreted is  
\begin{equation}	
M_{\rm acc}=1.3 \times 10^{43} 
\left (\frac{\rho_{\rm dm}}{0.3 {\rm GeV}/{\rm cm}^3} \right )  
\left (\frac{t}{{\rm Gyr}} \right )f~{\rm GeV}, 
\label{rate}  
\end{equation} 
where $\rho_{\rm dm}$ is the local dark matter density, and the ``efficiency'' factor $f=1$ if the WIMP-nucleon cross
section satisfies $\sigma >
10^{-45}{\rm cm}^2$, and $f=\sigma/(10^{-45}{\rm cm}^2)$ if $\sigma <
10^{-45}{\rm cm}^2$.

 One can easily check that $M_{\rm acc}$ can be larger than $M_{\rm crit}$ practically for all masses larger than $\sim 100$ keV.
 To form a black hole, satisfying the condition (\ref{m_crit}) is necessary but it is not sufficient. One should make sure that after the WIMPs have been captured, they slow down and thermalize with nuclear matter concentrating within a small thermal radius. Failing to satisfy this condition, even if the condition (\ref{m_crit}) is satisfied, does not necessarily lead to the formation of a black hole, since WIMPs would not be confined in a tiny region. The thermalization time scale has been estimated in~\cite{Goldman:1989nd} and \cite{Kouvaris:2010vv} 
 \begin{equation}
t_{\rm th} = 0.2 {\rm yr} \left({m\over {\rm TeV}} \right)^2
\left({\sigma\over 10^{-43}{\rm cm}^2}\right)^{-1}
\left({T\over 10^5{\rm K}}\right)^{-1}.
\end{equation} As one can observe, despite the Pauli blocked interactions between WIMPs and nucleons, unless they are very heavy, WIMPs thermalize in less than a year. Having thermalized with nuclear matter, WIMPs concentrate in the center of the star within a thermal radius that can be easily obtained by use of the virial theorem
\begin{equation}
r_{\rm th}= \left ( \frac{9kT_c}{8\pi G \rho_c m} \right )^{1/2}=220 {\rm cm} \left( \frac{{\rm GeV}}{m} \right )^{1/2} \left ( \frac{T_c}{10^5K} \right )^{1/2},
\end{equation}
where $k$ is the Boltzmann constant, $T_c$ is the temperature at the core of the star, and $\rho_c=5 \times 10^{38} {\rm GeV}/{\rm cm}^3$ is a typical value for the neutron star core density.

Once the WIMPs are thermalized and if sufficient number is accumulated in the star, there are two different events that take place, the time order of which depends on the WIMP mass. One is the self-gravitation of the WIMP sphere and the second is the formation of a Bose Einstein condensate (BEC). Self-gravitation takes place once the mass of the WIMP sphere inside the thermal radius becomes larger than the mass of the neutron star within the same radius. In other words, this happens once WIMPs start feeling strongly their own gravitational field. For this to happen the WIMP sphere should have a mass that satisfies
\begin{equation}
M_{\rm sg} > \frac{4}{3}\pi \rho_c r_{\rm th}^3 = 2.2 \times
10^{46}~{\rm GeV}
\left( {m\over {\rm GeV}} \right)^{-3/2}.
\label{eq:selfgrav}
\end{equation}
On the other hand, BEC formation takes place once the WIMP number density is
\begin{equation}
n_{\rm BEC} \simeq  4.7\times 10^{28} {\rm cm}^{-3} 
\left( {m\over {\rm GeV}} \right)^{3/2} 
\left (\frac{T_c}{10^5 {\rm K}} \right )^{3/2}.\label{BEC1}
\end{equation}
One can easily check that for WIMPs roughly lighter than 10 TeV, the accumulated WIMPs within $r_{\rm th}$ meet first the condition for BEC formation. We are going to study these two cases ($m<10$ TeV and $m>10$ TeV) separately since events unfold with different order. For WIMPs lighter than 10 TeV, one can estimate that the total number of WIMPs needed to form a BEC is $N_{\rm BEC} \simeq 2 \times 10^{36}$. Any accumulated WIMPs on top of this number goes directly to the ground state of the BEC state. The radius of the BEC state is 
\begin{equation}
r_{\rm BEC} = \left (
\frac{8\pi }{3} G \rho_c m^2\right )^{-1/4}\simeq 1.6 \times 10^{-4} \left
(\frac{{\rm GeV}}{m} \right )^{1/2}
{\rm cm}.
\label{condensed_ground}
\end{equation}
As it can be seen, $r_{\rm BEC}<<r_{\rm th}$ and therefore WIMPs in the ground state can become self-gravitating much faster than what Eq.~(\ref{eq:selfgrav}) predicts. In fact we can appreciate this if we  substitute  $r_{\rm th}$ by $r_{\rm BEC}$ in
Eq.~(\ref{eq:selfgrav}). This leads to the condition 
\begin{equation}
M > 8\times 10^{27}~{\rm GeV} 
\left( {m\over {\rm GeV}} \right)^{-3/2}.
\label{eq:}
\end{equation}
If Eqs.~(\ref{m_crit}),({\ref{BEC1}), and (\ref{eq:}) are satisfied, a black hole is going to be formed. Once the black hole is formed, its fate is determined by its initial mass $M_{\rm crit}$. One the one hand, the black hole is accreting dark matter and nuclear matter from the core of the star. This tends to increase the black hole mass. On the other hand, emission of photons and particles in general via Hawking radiation tends to reduce the mass of the black hole. The black hole mass evolution is determined by  
\begin{equation}
\frac{dM}{dt}=\frac{4\pi\rho_c G^2 M^2}{c_s^3}-\frac{f}{ G^2 M^2}, \label{eq:dMdt}
\end{equation}  where $c_s$ is the sound speed at the core of the star, and $f$ is a dimensionless number that in general depends on the number of particle species emitted and the rate of rotation of the black  hole. We have used a spherically symmetric Bondi accretion of matter into the black hole. By inspection of Eq.~(\ref{eq:dMdt}) it is apparent that there is a critical value of the black hole mass $M$ above which accretion always wins, while below, Hawking radiation reduces the mass of the black hole which in turn it increases even further the rate of Hawking radiation leading eventually to the evaporation of the black hole. This critical mass has been estimated if one considers only photons in~\cite{Kouvaris:2011fi} 
\begin{equation}
M > 5.7 \times 10^{36}~{\rm GeV}.
\label{eq:hawking}
\end{equation}
The mass becomes slightly larger~\cite{Fan:2012qy} if one includes also other species that can be emitted (e.g gravitons, neutrinos, quarks, leptons etc). Comparison of Eq.~(\ref{m_crit}) (with $\lambda=0$) to Eq.~(\ref{eq:hawking}) shows that WIMP masses larger than 16 GeV lead to black hole masses below the limit of Eq.~(\ref{eq:hawking}). This means that for masses larger than 16 GeV, black holes evaporate and their effect is to heat up the star as they evaporate. However this does not lead to a dramatic effect like the destruction of the star. This 16 GeV mass limit becomes slightly smaller if more Hawking radiation modes are included.

Finally there is one last constraint that should be satisfied. WIMP masses cannot be arbitrarily small because for small WIMP masses, after WIMPs have thermalized, those in the tail of the Maxwell-Boltzmann distribution have large enough velocities to escape from the star. This evaporation effect can be ignored for WIMP masses $m>2$ keV~\cite{Kouvaris:2011fi}.
If the accreted dark matter mass within a billion years $M_{\rm acc}$ is larger than $M_{\rm crit}$ of Eqs.~(\ref{m_crit}), and ({\ref{BEC1}), (\ref{eq:}), and (\ref{eq:hawking}) are satisfied, the WIMPs form a black hole that can destroy the star. There are some subtle issues regarding how fast the black hole consumes the star that have been addressed to some extend in~\cite{Kouvaris:2011fi}. The constraints on asymmetric bosonic dark matter are depicted in Fig.~1.
\begin{figure}[t]
  \centering
  \includegraphics[width=0.4\textwidth]{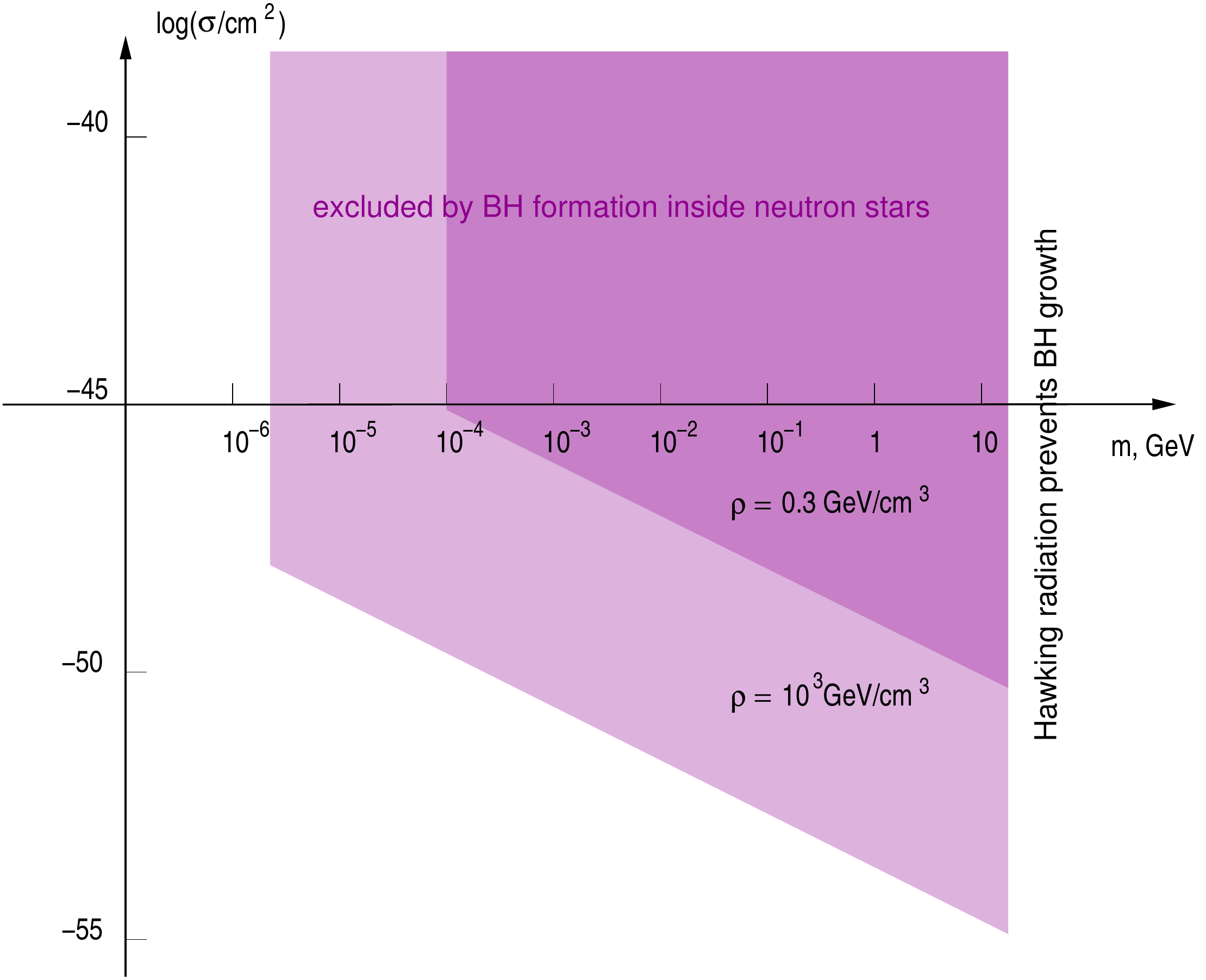}
  \caption{Exclusion regions of the asymmetric bosonic dark matter as a function of the WIMP mass and the WIMP-nucleon cross section for an isolated neutron star at local DM density  $\rho_{dm}= 0.3 {\rm GeV}/{\rm cm}^3$ (such as J0437-4715 and J0108-1431) and for a neutron star in the core of a globular cluster with $\rho_{dm} = 10^3 {\rm GeV}/{\rm cm}^3$.}
  \label{simp_fig}
 \end{figure}
 As it can be seen, depending on the WIMP-nucleon cross section, WIMP candidates from 100 keV up to roughly 16 GeV are severely constrained by the existence of nearby old neutron stars. The constrained region is bound at 100 keV due to the fact that below that mass accretion is not sufficient to acquire $M_{\rm crit}$ from Eq.~(\ref{m_crit}). These constraints can be enlarged down to 2 keV (the limit from WIMP evaporation we mentioned before) as long as we consider old neutron stars in globular clusters with $\rho_{\rm dm} \gtrsim 30~{\rm GeV}/{\rm cm}^3$.
 
Now we can consider the case where the WIMP mass is larger than 10 TeV and therefore self-gravitation of the WIMP sphere happens before BEC formation. As we mentioned above,  black holes of critical mass~(\ref{m_crit}) with WIMP masses roughly larger than $\sim 16$ GeV, do not survive due to Hawking radiation. Therefore one should expect that black holes of $M_{\rm crit}$ (of Eq.~(\ref{m_crit})) formed out of 10 TeV WIMPs (or heavier) would evaporate quite fast. However, since self-gravitation takes place before BEC, and the self-gravitating mass of Eq.~(\ref{eq:selfgrav}) for $m>10$ TeV is much larger than the crucial mass for the survival of the black hole of Eq.~(\ref{eq:hawking}), there were speculations in the literature~\cite{McDermott:2011jp,Guver:2012ba,Fan:2012qy} 
that constraints can be imposed also for $m>10$ TeV. The claim was that instead of forming a black hole of $M_{\rm crit}$ that is below the surviving threshold for Hawking radiation, a much larger black hole coming from the collapse of the self-gravitating WIMP sphere $M_{\rm sg}$ forms, that due to its larger mass can grow and destroy the star, thus imposing constraints on this part of the parameter space of asymmetric bosonic dark matter. However we review here the argument that was put forward in~\cite{Kouvaris:2012dz} that demonstrates that the formation of smaller (non-surviving) black holes of mass $M_{\rm crit}$ is unavoidable and therefore the $M_{\rm sg}$ instead of collapsing to a single large black hole, it forms a series of black holes of $M_{\rm crit}$ that evaporate one after the other, thus resulting to no constraint for WIMP masses with $m>10$ TeV.

In order for the WIMP sphere to collapse, the whole mass should be confined within the Schwarzschild radius $r_s=2GM$ of the black hole. The density of WIMPs just before forming the black hole would be $n_{\rm BH} \sim 3 (32\pi G^3 M_{\rm sg}^2m)^{-1} \sim 10^{74}~ {\rm cm}^{-3}({\rm GeV}/m) (M_{\rm sg}/10^{40}{\rm GeV})^{-2}$.  It is easy to see that this density is higher from the density required for BEC formation of Eq.~(\ref{BEC1}). This means that unless the WIMP sphere collapses violently and rapidly, it should pass from a density where BEC is formed. 
As the self-gravitating WIMP sphere of mass $M_{\rm sg}$ contracts, at some point it will reach the density where BEC is formed. Any further contraction of the WIMP sphere will not lead to an increase in the density of the sphere. The density remains that of BEC. The formation of BEC happens on time
scales of order \cite{Stoof:1992zz}
$t_{\rm BEC} \sim \hbar/k_BT  \sim10^{-16}{\rm s}$, i.e. practically instantaneously. Further shrinking of the WIMP
sphere results in increasing the mass of the condensate rather than
the density of non-condensed WIMPs. This process happens at a time scale which is determined by the cooling time of the WIMP sphere as discussed below. As we shall show, this cooling time is the relevant time scale for the BEC formation.
 As in the previous case, the ground state will start being populated with WIMPs which at some point will become self-gravitating themselves. This of course will happen not when Eq.~(\ref{eq:}) is satisfied. Eq.~(\ref{eq:}) was derived as the WIMP ground state becomes denser than the surrounding nuclear matter (since the dark matter that is not in the ground state of the BEC is less dense). Here, the condition is that the density of the ground state of the BEC should be larger than the density of the surrounding dark matter (that is already denser than the nuclear matter at this point). The condition reads
\begin{equation}
M_{\rm BEC,\,sg} = {4\pi\over 3} n_{\rm BEC}m r_{\rm BEC}^3 
= 9.6 \times 10^{21}{\rm GeV} \left({m\over10 {\rm TeV}}\right)^{-7/8}.
\end{equation}
Once the BEC ground state obtains this mass, the ground state starts collapsing within the collapsing WIMP sphere. Any contraction of the WIMP sphere does not change the density of the sphere but only the density of the ground state. $M_{\rm BEC,\,sg}$ is smaller than $M_{\rm crit}$ and therefore the BEC ground state cannot form a black hole yet. However as the ground state gets populated at some point it reaches the point where its mass is $M_{\rm crit}$ and this leads to the formation of a black hole of mass $M_{\rm crit}$ and not $M_{\rm sg}$. The evaporation time for such a black hole of $M_{\rm crit}$ (composed of WIMPs of mass $m$) is
\begin{equation}
\tau = 5\times 10^3 {\rm s} \left({10 {\rm TeV}\over m }\right)^3. \label{evap}
\end{equation}
The only way that such a black hole can be maintained in life is by adding fast new matter inside in a rate that is higher than the Hawking radiation. As we showed earlier accretion of nuclear matter from the core of the star is not sufficient for $m$ larger than roughly 16 GeV. Accretion of dark matter from the rest of the WIMP sphere is also not sufficient. The
accretion of non-interacting collisionless particles  in the
non-relativistic limit onto a black hole is given by~\cite{Nov}
\begin{equation}
F=\frac{16 \pi G^2 M_{\rm BH}^2 \rho_{\rm dm}}{v_\infty},
\end{equation}
where $\rho_{\rm dm}$ is $mn_{\rm BEC}$, and $v_{\infty}$ is the average WIMP
velocity far away from the black hole. Using the virial theorem, we
take $v_\infty= \sqrt{GM/r}$, $M$ being the total mass of the WIMP
sphere and $r$ is the radius of the WIMP sphere. For the first black hole $M=M_{\rm sg}$ and $r=r_0$ where $r_0$ is the radius of the WIMP sphere when it reaches the BEC density i.e.
\begin{equation}
r_0= \left ( \frac{3M_{{\rm sg}}}{4 \pi m n_{\rm BEC}} \right)^{1/3}
=2.2\, {\rm cm} \left (\frac{10 {\rm TeV}}{m} \right )^{4/3} . 
\label{eq:r0}
\end{equation}
  For WIMP sphere masses ranging from Eq.~(\ref{m_crit}) to
Eq.~(\ref{eq:selfgrav}) the Hawking radiation dominates
overwhelmingly over the WIMP accretion despite the large WIMP density, and
therefore the black hole evaporation time is given accurately by
Eq.~(\ref{evap}). 

There are another two ways where the black hole can be maintained in life i.e. if the WIMP capture time scale and/or the cooling time scale are smaller than $\tau$. The capture time scale is defined as the time it takes for the star to capture a number of WIMPs with a total mass of $M_{\rm crit}$. If this time is smaller than $\tau$, a new black hole of $M_{\rm crit}$ will form before the extinction of the previous one. Therefore it is possible that gradually the mass of the black hole could increase after successive black holes form before previous ones have evaporated. However, using Eqs.~(\ref{m_crit}) and (\ref{rate}) we get the capture time scale
\begin{equation}
t_{\rm cap} = 7.6\times 10^4 {\rm s} \left({m\over10 {\rm
    TeV}}\right)^{-1}
\left({ \rho_{\rm dm} \over 10^3 {\rm GeV/cm^3}}\right)^{-1}, 
\end{equation}
which is larger than $\tau$ for $m>10$ TeV even for neutron stars located in high dark matter density globular cluster cores. Capture rates of new WIMPs cannot replenish the black hole mass rate loss to Hawking radiation. 

Finally one should estimate the cooling time scale for the WIMP sphere. Once self-gravitating, the WIMP sphere does not collapse to a black hole just because WIMPs still have a lot of energy. This energy can be lost via collisions with nucleons. This way the WIMPs lower their energy and the WIMP sphere contracts. Let us assume that a black hole of mass $M_{\rm crit}$ has been formed and it is going to evaporate in $\tau$. Once formed the black hole is doomed  to evaporate unless the WIMP sphere contracts fast enough, so it replenishes the BEC ground state with another $M_{\rm crit}$ mass that collapses and forms a second black hole, before  the first one has evaporated. In that case, the mass of the black hole would be reinforced with new mass making possible for the black hole to grow. As we shall show the time scale it takes for the WIMP sphere to populate again the BEC ground state is larger than the evaporation time and therefore the black hole cannot survive. The energy that the WIMP sphere has to lose in order to admit $M_{\rm crit}$ in the BEC ground state is
\begin{equation}
\delta E = {1\over 2} {GMM_{\rm crit} \over r}, 
\label{eq:energygain}
\end{equation}
where $M$ is the total WIMP mass and $r$ is the radius of the WIMP-sphere. For the first black hole 
 $M=M_{\rm sg}$ and $r=r_0$. This energy must be lost by WIMP-nucleon collisions. The time it takes on average for a WIMP-nucleon collision (taking into account the Pauli blocking effect for the degenerate nucleons) is 
\begin{equation}
\tau_{\rm col} = {1\over n \sigma v}\frac{4 p_F}{3m_N v}
= {2 p_F m\over 3\rho_c\sigma \epsilon},
\label{eq:tau_collision}
\end{equation}
where $p_F\simeq 1$ GeV is the nucleon Fermi momentum at the core of the star, $n$ the number density of nucleons, $m_N$ the nucleon mass, $\sigma$ the WIMP-nucleon cross section, and $\epsilon$ is the kinetic energy of the WIMP. $\epsilon$ can be estimated by use of the virial theorem
\begin{equation}
\epsilon \sim {GMm\over 2 r_0}. 
\label{eq:WIMPenergy}
\end{equation}
One should keep in mind that each WIMP-nucleon collision leads to a loss of energy $\delta \epsilon = 2(m_N/m)\epsilon$.Therefore in order for $N=M/m$ WIMPs to lose the excess $\delta E$ energy colliding in $\tau_{\rm col}$ time scales with nucleons losing each time on average energy $\delta \epsilon$, it takes 
\begin{equation}
t_{\rm cool} =\tau_{\rm col}{ \delta E \over N \delta\epsilon }  =\tau_{\rm col} {m \delta E \over M \delta\epsilon } 
= {4 \over 3\pi}{ p_F\over m_N} 
{ r_0 M_{\rm Pl}^4 \over \rho_c \sigma M^2} .
\label{eq:t_cool}
\end{equation}
One can see that this time is shorter for larger 
mass $M$.  Substituting $M=M_{\rm sg}$ into Eq.~(\ref{eq:t_cool})
one gets 
\[
t_{\rm cool} \simeq 
1.5\times 10^3 {\rm s} 
\]
\begin{equation}
\times
\left({m\over10 {\rm  TeV}}\right)^{5/3}
\left({T\over 10^5 {\rm  K}}\right)^{-3}
\left({\sigma\over 10^{-43}\,{\rm cm}^2}\right)^{-1}.
\label{eq:t_cool_fin}
\end{equation}
This time is shorter by a factor of a few than the black hole evaporation time of
Eq.~(\ref{evap}). Note, however, the strong dependence of both quantities on
the WIMP mass $m$. Already for masses $m>13$ TeV  the black hole evaporation
time becomes shorter. The
cooling time derived above refers to the formation of the first black
hole. Subsequent black holes
require progressively longer times. This is easily seen from
Eq.~(\ref{eq:t_cool}) because $t_{\rm cool}$ scales as $1/M^{5/3}$
(recall that $r_0$ scales as $M^{1/3}$ from Eq.~(\ref{eq:r0})). Thus,
the more black holes have formed and evaporated, the smaller is the remaining
WIMP mass $M$ and the longer the time needed to form the next one.  
\section{Conclusions}
We reviewed the current status of constraints on non-self-interacting asymmetric bosonic dark matter based on observations of old neutron stars. We showed that severe constraints can be imposed for these candidates from the keV to the GeV range, excluding all candidates with WIMP-nucleon cross sections like the ones favorited by the DAMA experiment within this mass range.
We also reviewed the arguments why these constraints cannot be extended to candidates with masses in the TeV range or higher.
   
  \end{document}